\documentclass[a4paper,10pt]{article}
\usepackage{graphicx}
\usepackage{epstopdf}

\usepackage{dcolumn}

\usepackage{multirow}

\usepackage{amsfonts,amsmath,amssymb}
\usepackage{hyperref}

\usepackage{epsfig,multicol}

\newcommand\fverb{\setbox\pippobox=\hbox\bgroup\verb}
\newcommand\fverbit{\egroup\item[\fbox{\unhbox\pippobox}]}

\newbox\pippobox

\begin{document}

\title{\bf Chameleon Field Dynamics During Inflation}
\author{Nasim Saba\thanks{Electronic address: n$\_$saba@sbu.ac.ir}\, and\,
        Mehrdad Farhoudi\thanks{Electronic address: m-farhoudi@sbu.ac.ir}\,\,
\\
\small Department of Physics, Shahid Beheshti University, G.C.,
       Evin, Tehran 19839, Iran}
\date{\small December 13, 2017}
\maketitle

\begin{abstract}
\noindent
 By studying the chameleon model during inflation, we
investigate whether it can be a successful inflationary model,
wherein we employ the common typical potential usually used in the
literature. Thus, in the context of the slow--roll approximations,
we obtain the number of e--folding for the model to verify the
ability of resolving the problems of standard big bang cosmology.
Meanwhile, we apply the constraints on the form of the chosen
potential and also on the equation of state parameter coupled to
the scalar field. However, the results of the present analysis
show that there is~not much chance of having the chameleonic
inflation. Hence, we suggest that if through some mechanism the
chameleon model can be reduced to the standard inflationary model,
then it may cover the whole era of the universe from the inflation
up to the late time.
\end{abstract}

\bigskip

\section{Introduction}
\indent

The recent accelerated expansion of the universe has been reported
by the various observed cosmological data such as the luminosity
redshift relation for the supernovae type Ia
(SNIa)~\cite{1}--\cite{2}, the large scale structure
formation~\cite{lss}, the baryon acoustic oscillation
(BAO)~\cite{benitez,parkinson}, and the cosmic microwave
background (CMB) temperature anisotropies measured by some
experiments such as COBE~\cite{cobe},
WMAP~\cite{spergel}--\cite{wmap2} and
Planck~\cite{p13}--\cite{p152}. Numerous attempts have been
proposed to present a theoretical explanation to this mysterious
acceleration of the universe, which could have arisen from a dark
energy component or being due to departure of gravity from general
relativity on cosmological scales, see, e.g.,
Refs.~\cite{nojiri}--\cite{ZareFarhoudi}. In the former mechanism,
dark energy is ``some kind of matter" with a negative pressure
that is supposed to be responsible for the accelerated expansion
of the universe. However, the important point is that the
parameters entering each mechanism must satisfy both the current
astronomical observations and the laboratory experiments.

Amongst different models for dark energy, the cosmological
constant is the simplest one in which its energy density is
constant. As this model has some difficulties, such as the
cosmological constant problem~\cite{weincc}--\cite{bernard} and
the coincidence problem~\cite{sahni,carroll2,bernard}, scalar
fields have been introduced as dark energy component with a
dynamical equation of state, see, e.g.,
Refs.~\cite{peebles}--\cite{bamba}. Essentially, scalar fields
have long history in physics, and more recently, they have played
an important role in both cosmology and particle physics. In fact,
it is believed that our universe consists of some scalar fields in
addition to the matter fields. In this context, the scalar--tensor
theory of gravitation recently becomes one of the most popular
alternatives$^{1}$\footnotetext[1]{For a review on alternative
theories of gravitation, see, e.g., Ref.~\cite{Farhoudi-2006} and
references therein.}\
 to the Einstein gravitational theory, see, e.g.,
Refs.~\cite{capozziello,fujji}--\cite{Farajollahi-2010}. In
particular, quintessence is a more general dynamical model in
which the energy source of the universe, unlike the cosmological
constant, varies in space and time~\cite{cald}--\cite{stein}.

On the other hand, the higher dimensional gravities, e.g.~the
string theory and supergravity, predict some massless scalar
fields that couple directly to the matter with gravitational
strength, see, e.g., Refs.~\cite{brustein,bin}. These theories may
motivate ones to investigate the scalar--tensor theory wherein the
scalar field couples to the matter with the gravitational
strength. However, due to such coupling, a fifth force and also
large violation of the equivalence principle (EP) should be
detected~\cite{fi,euc} in contrast to the results of the solar
system tests of gravity~\cite{iess}--\cite{dec}. Thus, if one
considers such coupling to the matter for the quintessence field,
then some mechanisms must effectively screen those resulted forces
locally and also prevent the EP--violation. In fact, the notion of
screening mechanisms is how a scalar field can act as dark energy
on cosmological scales while being shielded in the regions of high
density, such as on the earth (see, e.g.,
Refs.~\cite{screen1,screen2} for a review on this issue). There
are usually two main ways for having a suitable scalar field model
while avoiding the EP--violation:
\begin{itemize}
\item Such a scalar field must usually be very light and its coupling to the
      matter should be tuned to extremely small values~\cite{uzan}--\cite{symetron}.
\item The scalar field can couple to the matter fields with the gravitational
      strength and then, acquires a mass depending on the background
      matter density of the environment which leads its interaction
      to be effectively short--ranged~\cite{weltman1}--\cite{Burrage2017}.
\end{itemize}

In this regard, the chameleon model proposed in
Refs.~\cite{weltman1,weltman2} invokes a screening mechanism based
on the second approach remaining consistent with the tests of
gravity on the terrestrial and the solar system scales. Actually,
in this type of chameleon cosmology, the corresponding scalar
field couples to an ambient matter field through a conformal
factor, relating the Jordan and Einstein frames, with the
gravitational strength, and consequently, its mass is no longer
fixed but depends on the environmental situations. In the dense
environments, such as on the earth, the chameleon field acquires a
large mass that makes its effects short--ranged, and hence,
becomes invisible to be searched for the EP--violation and fifth
force in the current experimental and observational tests. Such a
property of the chameleon field is of great deal of importance. On
the other hand, it is very light in diluted matter situations,
such as the cosmological scales, and thus, the chameleon field may
play the role of dark energy causing the cosmic late time
acceleration.

However, under a very general conditions, there were proved two
theorems by which it has been claimed that they limit a
cosmological impact of the chameleon field~\cite{wang2,khoury}.
Actually, it was shown that at the present cosmological density
the Compton wavelength of the chameleon field should be of the
order of $1$ Mpc, and the conformal factor during the last Hubble
time is almost a practically constant value. Their results imply
that chameleon--like scalar fields have a negligible effect on
density perturbations on linear
scales\rlap,$^{1}$\footnotetext[1]{It is worth mentioning that
this is~not the case on non--linear scales since this is an active
area of the chameleon research.}\
 and may~not account for the
observed cosmic acceleration except as some form of dark energy.
Nevertheless, it has been indicated~\cite{ivanov} that even if the
cosmological factor being in principle constant during the last
Hubble time, this will~not prevent for the chameleon field to be
responsible for a late--time acceleration of the universe
expansion. Indeed, the conformal factor describing the interaction
of the chameleon field with an ambient matter may lead to
deviations of the matter and radiation densities in the universe
from their canonical forms $1/a^3$ and $ 1/a^4$, where $a$ is the
cosmological factor describing a dynamics of the universe
expansion. However, one may assert that these results
in~\cite{ivanov} are model--independent, since the shape of the
potential of the self--interaction of the chameleon field was~not
specified. In addition, it has been shown~\cite{ivanov2} that the
cosmological constant and dark energy density might be induced by
torsion in the Einstein--Cartan gravitational theory. Actually, as
torsion is a natural geometrical quantity additional to the metric
tensor, the analysis of the nature of the cosmological constant
performed in~\cite{ivanov2} provides a robust geometrical
background for the origin of the cosmological constant and dark
energy density.

In the present work, we propose to study the chameleonic behavior
during another important acceleration era, namely the inflation
which is an extremely rapid expansion that resolves some important
problems of the standard cosmological model such as the flatness,
horizon and monopole~\cite{gu}--\cite{fara}. By this paper, we aim
to investigate whether the chameleon model is a successful model
during the inflation, and what would be the ambient matter of the
universe during the inflation that the chameleon field is coupled
to. In this case, one may claim that the chameleon field, as a
single scalar field, being responsible for the acceleration of the
universe both at the very early and at the late time. To perform
this task, we consider a coupling between the chameleon scalar
field and an unknown matter scalar field with the equation of
state parameter $w$. In addition, to probe the behavior of the
model in the inflationary era, we expect to specify the value of
$w$ via the constraints during the analysis of the results in
order to figure out the type of this unknown coupled matter field.

Moreover, the possibility of describing the chameleon and the
inflaton by one single scalar field has also been investigated in
Ref.~\cite{ch} where the scenario is based on a modified
supersymmetric potential introduced by Kachru, Kallosh, Linde and
Trivedi (KKLT)~\cite{kklt}. Indeed, by using this modified
construction of KKLT potential, they have attempted to embed the
chameleon scenario within the string compactifications, where it
has been shown that the volume modulus of the compactification can
act as a chameleon field. The late time investigation has been
presented in Ref.~\cite{h1}, while Ref.~\cite{ch} describes the
scenario during the inflation wherein it has been presented that
in order to cover the cosmology of both the late time and the very
early universe, there exists a superpotential consisting of two
pieces, which one drives the inflation in the very early universe,
and the other one is responsible for the chameleon screening at
the late time.

The work is organized as follows. In the next section, we
introduce the model and obtain the field equations of motion by
taking the variation of the action. In Sect.~$3$, we investigate
the model during the inflation by imposing the slow--roll
approximations. Also, we compute the value of e--folding number of
the model for the common typical potential usually used in the
context of the chameleon theory in the literature, and then, set
the constraints to pin down the free parameters of the model.
Finally, we conclude the work in Sect.~$4$ with the summary of our
results.

\section{Chameleon Scalar Field Model}\label{Sec2}
\indent

We start with the following known action that governs the dynamics
of the chameleon scalar field model in 4--dimensions, i.e.
\begin{eqnarray}
S\!\!\!&=\!\!\!&\!\!\int{\rm
d}^{4}x\sqrt{-g}\left[\frac{M^{2}_{\rm
Pl}R}{2}\!-\!\frac{1}{2}\partial_{\mu}\phi \partial^{\mu}\phi-V(\phi)\right]\nonumber
\\
 &&+\sum_{i}\int{\rm d}^{4}x\sqrt{-{\tilde g}^{(i)}}{L_{\rm m}^{(i)}}\left({\psi}^{(i)},{\tilde g}^{(i)}_{\mu\nu}\right),
\end{eqnarray}
where $R$ is the Ricci scalar constructed from the metric
$g_{\mu\nu}$ with signature $(-,+,+,+)$, $g$ is the determinant of
the metric, $M_{\rm Pl}\equiv(8\pi G)^{-1/2}\approx 10^{27}eV$ is
the reduced Planck mass (we use the units in which $c=1=\hbar$)
and the lowercase Greek indices run from zero to three. Also,
$\phi$ is a scalar field, $V(\phi)$~is a self--interacting
potential, $L_{\rm m}^{(i)}$'s~are the Lagrangians of the matter
fields, $\psi^{(i)}$'s~are various matter scalar fields, $\tilde
g_{\mu\nu}^{(i)}$'s are the matter field metrics that are
conformally related to the Einstein frame metric $g_{\mu\nu}$ via
\begin{equation}
{\tilde g}^{(i)}_{\mu \nu}={e}^{ {2\frac {{\beta} _{i}\phi}{M_{\rm Pl}}}}{g}_{\mu \nu},
\end{equation}
where $\beta_{i}$'s~are dimensionless constants representing
different non--minimal coupling constant between the scalar field
and each matter species. However, in our analysis, we just focus
on a single matter component, and henceforth, we drop the index
$i$. We also consider the universe to be described by the
spatially flat Friedmann--Lema\^{\i}tre--Robertson--Walker ({\bf
FLRW}) metric in the Einstein frame as
\begin{equation}\label{metric}
ds^{2}=-dt^{2}+a^{2}(t)\left(dx^{2}+dy^{2}+dz^{2}\right),
\end{equation}
where $t$ is the cosmic time. Obviously, the corresponding
metric in the Jordan frame is
\begin{equation}\label{metric2}
d\tilde{s}^{2}=-{e}^{ {2\frac {{\beta}\phi}{M_{\rm Pl}}}}dt^{2}+
\tilde{a}^{2}(t)\left(dx^{2}+dy^{2}+dz^{2}\right),
\end{equation}
where $\tilde{a}(t)$ is the scale factor in the Jordan frame, i.e.
$\tilde{a}(t)\equiv a(t)\exp({\beta\phi/M_{\rm Pl}})$.

Varying the action with respect to the scalar field yields the field equation of motion
\begin{equation}\label{box}
\Box\phi=V'(\phi)-\frac{{\beta}}{M_{\rm
Pl}}e^{4\frac{{\beta}\phi}{{M}_{\rm Pl}}}{\tilde
g}^{\mu\nu}{\tilde T}_{\mu\nu},
\end{equation}
where $\Box\equiv\nabla^{\alpha}\nabla_{\alpha}$ corresponding to
the metric $g_{\mu\nu}$, the prime is the derivative with respect
to the argument and $\tilde T_{\mu\nu}=-(2/\sqrt{-\tilde
{g}})\delta (\sqrt{-\tilde {g}}L_{\rm m})/\delta {\tilde
g}^{\mu\nu}$ is the energy--momentum tensor that is conserved in
the Jordan frame, i.e. $\tilde\nabla_{\mu} {\tilde T^{\mu\nu}}=0$.
We assume an unknown matter field as a perfect fluid with the
equation of state $\tilde p=w\tilde \rho$ that, in the FLRW
background, one has
\begin{equation}
{\tilde g}^{\mu\nu}\tilde T_{\mu\nu}=-(1-3w)\tilde \rho,
\end{equation}
where $\tilde\rho$ is the matter density in the Jordan frame. As
$\tilde\rho$ is~not conserved in the Einstein frame, we propose to
have a conserved matter density with the same equation of state,
which is independent of $\phi$ and obeys the following relation in
the Einstein frame
\begin{equation}\label{cont1}
\rho\propto a^{-3(1+w)}.
\end{equation}
In this respect, as the continuity equation for $\tilde\rho$ in the Jordan frame is
\begin{equation}\label{cont2}
\dot{\tilde{\rho}}+3\frac{\dot{\tilde{a}}}{\tilde{a}}(1+w)\tilde{\rho}=0,
\end{equation}
that yields $\left(\tilde{a}^{3(1+w)}\tilde{\rho}\right)_{,0}=0$,
hence in order to have relation (\ref{cont1}), one can define the
conserved matter density in the Einstein frame as
\begin{equation}\label{rhoo}
\rho\equiv e^{3(1+w)\frac{{\beta}\phi}{M_{\rm Pl}}}\tilde \rho.
\end{equation}
Therefore, equation (\ref{box}) reads$^{1}$\footnotetext[1]{Note
that, through our analysis, the matter field has been coupled to
the Einstein frame metric.}
\begin{equation}
\Box\phi=V'(\phi)+(1-3w)\frac{\beta}{{M}_{\rm
Pl}}\rho e^{(1-3w)\frac{{\beta}\phi}{M_{\rm Pl}}},
\end{equation}
and thus, the dynamic of the scalar field is actually governed by
an effective potential, i.e.
\begin{equation}\label{field}
\Box\phi=V'_{\rm eff}(\phi),
\end{equation}
where
\begin{equation}\label{eff}
{V}_{\rm eff}(\phi)\equiv V(\phi)+\rho(t)\,
e^{(1-3w)\frac{\beta\phi}{M_{\rm Pl}}}.
\end{equation}
As it is obvious, the effective potential depends on the
background matter density $\rho$ of the environment. Consequently,
the value of $\phi$ at the minimum of $V_{\rm eff}$ and the mass
fluctuation about the minimum\rlap,$^{2}$\footnotetext[2]{The mass
of the scalar field can analogously be defined as
$m_{\phi}\equiv\sqrt{V''_{\rm eff}(\phi_{\rm min})}$.}
 depend on
the matter density. For instance, if one considers $V(\phi)$ as a
decreasing function of $\phi$, while $\beta>0$ and $1-3w>0$, the
minimum of the effective potential decreases by increasing $\rho$,
hence, the mass of the scalar field will increase in a way that
the chameleon field can be hidden from the local experiments.

By the FLRW metric (\ref{metric}), the field equation
(\ref{field}) gives the corresponding Klein--Gordon equation
\begin{equation}\label{phi}
\ddot{\phi}+3H\dot\phi+V'_{\rm eff}(\phi) = 0,
\end{equation}
where $H(t)\equiv\dot{a}/a$ is the Hubble expansion rate of the
universe, dot denotes the derivative with respect to the cosmic
time, and the scalar field is only a function of the time due to
the homogeneity and isotropy. In addition, one can obtain the
Friedmann--like equations for the model by the variation of the
action with respect to the metric $g_{\mu\nu}$ in the context of
the perfect fluid as
\begin{equation}\label{friedman}
3M^{2}_{\rm Pl}H^{2}=\frac{1}{2}\dot\phi ^{2}+V_{\rm eff}(\phi)
\end{equation}
and
\begin{eqnarray}\label{fried2}
2\frac{\ddot{a}}{a}+H^{2}=-\frac{1}{M^{2}_{\rm Pl}}\left[\frac{1}{2}\dot{\phi}^{2}-V(\phi)
+w\rho e^{\left(1-3w\right)\frac{\beta\phi}{M_{\rm Pl}}}\right].
\end{eqnarray}
And in turn, the time derivative of the Hubble parameter, that
will be needed later on, is
\begin{equation}\label{hdot1}
M^{2}_{\rm
Pl}\dot{H}=-\frac{1}{2}\dot{\phi}^2-\frac{1}{2}(1+w)\rho
e^{\left(1-3w\right)\frac{\beta\phi}{M_{\rm Pl}}}.
\end{equation}
Now, in the following section, we mainly consider these equations
of motion in the inflationary era.

\section{Chameleon During Inflation}
\indent

In this section, by applying the common typical chameleonic
potential, we propose to investigate the chameleon model during
the inflation while considering a non--minimal coupling between
the chameleon field and an unknown matter scalar field with the
equation of state parameter $w$ which one may expect to be
specified via the constraints during the analysis. It is
remarkable that the chameleon field does~not couple to the
radiation (with $w=1/3$) as the trace of its energy--momentum
tensor is zero.

In this context, a potential, that has mostly been considered in
the literature, is in the form of inverse
power--law~\cite{weltman1}--\cite{brax}, although a $\phi^{4}$
potential has also been investigated~\cite{gubser}. To cover all
these kind of potentials, we consider it to be in general form
\begin{equation}\label{5}
V(\phi)=\lambda M^4\left(M/\phi\right)^{n},
\end{equation}
where $\lambda>0$ is a constant, $M$ is some mass scale and $n$ is
a positive or negative integer constant (however, see the point
below relation (\ref{phii})). Furthermore, when $n\neq-4$, $M$ can
be scaled such that, without loss of generality, one can set
$\lambda$ equals to unity (although, we have kept it), whereas for
$n=-4$, $M$ drops out and the $\phi^4$ theory is
resulted~\cite{mota1,mota2,t8,Burrage2017}.

On the other hand, a successful inflationary model should resolve
the puzzling issues of the standard big bang cosmology such as the
flatness, horizon and monopole problems.  In fact, an extremely
rapid expansion would cause the homogeneity and isotropy of the
universe at large scales. Amongst the mentioned difficulties, the
horizon problem is more important than the others, for its
resolution makes the other problems being solved
automatically~\cite{wein2}. Thus, to check whether the model
solves the mentioned problems, one should find the e--folding
number, where it is believed that a viable inflationary model
requires that the universe inflates by at least $50$ to $60$
times~\cite{karami}, or even more (i.e., nearly $70$ times or
higher). In addition, we assume that the value of the effective
potential being {\it almost} constant, i.e. almost being
equivalent to a quasi de Sitter expansion, then, we can obtain the
number of e--folding by imposing the slow--roll conditions for the
model as the most inflationary models are built upon the
slow--roll approximations.

In this respect, we use the Hubble slow--roll parameters defined
as$^{1}$\footnotetext[1]{These are related to the Hamilton--Jacobi
slow--roll parameters~\cite{liddle}; and meanwhile, the second
parameter may also be defined as ${\dot \epsilon}/(H\epsilon)$ for
$(\eta -\epsilon )$. In addition, in the literature, different
sets of slow--roll parameters have been used, however, any
inflationary model can be described by the evolution of one of the
relevant sets of parameters, see, e.g.,
Refs.~\cite{Liddle94,Makarov}.}
\begin{equation}\label{approx}
\epsilon \equiv  -\frac{\dot H}{H^{2}}  \qquad  \qquad   {\rm and}
\qquad   \qquad    \eta\equiv
-\frac{|\ddot{\phi}|}{H|\dot{\phi}|}.
\end{equation}
Now, during the inflation, to have an accelerated expansion of the
scale factor (i.e., $\ddot{a}>0$) and a sufficiently long enough
inflation (in order to solve the horizon problem), the slow--roll
parameters must be very smaller than unity, i.e., $\epsilon\ll1$
and $|\eta|\ll1$~\cite{liddle,wein2}, that lead to the following
two slow--roll conditions
\begin{equation}\label{ap}
\dot\phi^2 \ll |V-\frac{1+3w}{2}\rho
e^{(1-3w)\frac{\beta\phi}{M_{\rm Pl}}}|\qquad \qquad {\rm and}
\qquad \qquad |\ddot{\phi}|\ll H|\dot{\phi}|,
\end{equation}
where the first one is obtained by substituting equations
(\ref{friedman}) and (\ref{hdot1}) into definition $\epsilon$, and
guaranties slowly rolling of the scalar field during the
inflation. Note that, this condition is different from the
corresponding one in the standard model due to the non--minimal
coupling term. Under these slow--roll conditions, equations
(\ref{phi}), (\ref{friedman}) and (\ref{fried2}) reduce to
\begin{eqnarray}\label{appro}
\dot\phi\approx-\frac{1}{3H}V'_{\rm eff}(\phi),
\end{eqnarray}
\begin{eqnarray}\label{approo}
H^{2}\approx\frac{1}{3M^{2}_{\rm Pl}}V_{\rm eff}(\phi)
\end{eqnarray}
and
\begin{eqnarray}\label{approoo}
2\frac{\ddot{a}}{a}+H^{2}\approx -\frac{1}{M^{2}_{\rm
Pl}}\left[-V(\phi)+w\rho e^{\left(1-3w\right)
\frac{\beta\phi}{M_{\rm Pl}}}\right].
\end{eqnarray}
Thus, the slow--roll parameters (\ref{approx}) themselves read
\begin{equation}\label{epsi}
\epsilon\approx \epsilon_{1}+\frac{3(1+w)\rho e^{\left(1-3w\right)
\frac{\beta\phi}{M_{\rm Pl}}}}{2V_{\rm eff}(\phi)}
\end{equation}
and
\begin{equation}\label{ett}
\eta\approx \eta_{1}-\epsilon+\frac{3(1+w)(1-3w)\beta\rho
e^{\left(1-3w\right) \frac{\beta\phi}{M_{\rm Pl}}}}{M_{\rm
Pl}V'_{\rm eff}(\phi)},
\end{equation}
where the first relation is obtained using equations
(\ref{hdot1}), (\ref{appro}) and (\ref{approo}), and the second
one yields through taking the time derivative of equation
(\ref{appro}), then employing equations (\ref{appro}) and
(\ref{approo}) wherein considering the matter density conservation
in the Einstein frame, i.e. $\dot{\rho}=-3H(1+w)\rho$. Also, in
the same analogy with the standard model~\cite{liddle,wein2}, we
have defined $\epsilon_{1}$ and $\eta_{1}$ as
\begin{equation}\label{epsilon}
\epsilon_{1} \equiv \frac{M^2_{\rm Pl}}{2}\left(\frac{V'_{\rm
eff}(\phi)}{V_{\rm eff}(\phi)}\right)^{2}
\end{equation}
and
\begin{equation}\label{epsilon11}
\eta_{1} \equiv M^{2}_{\rm Pl}\frac {V''_{\rm eff}(\phi)}{V_{\rm eff}(\phi)}.
\end{equation}
However, due to the presence of the matter field, the slow--roll
parameters are different from the standard model. Now,
substituting potential (\ref{5}) into equation (\ref{epsi}) leads
to
\begin{eqnarray}\label{ep}
\epsilon \approx \frac{M^{2}_{\rm
Pl}}{2}\left[\frac{-n\frac{\lambda M^{4+n}}{\phi^{n+1}}+
\frac{\left(1-3w\right)\beta}{M_{\rm Pl}}\rho
e^{\frac{\left(1-3w\right)\beta\phi}{M_{\rm Pl}}}}{\frac{\lambda
M^{4+n}}{\phi^{n}}+\rho
e^{\frac{\left(1-3w\right)\beta\phi}{M_{\rm Pl}}}}\right]^{2}
+\frac{3}{2}\left[\frac{(1+w)\rho
e^{\left(1-3w\right)\frac{\beta\phi}{M_{\rm Pl}}}}{\frac{\lambda
M^{4+n}}{\phi^{n}}+\rho
e^{\frac{\left(1-3w\right)\beta\phi}{M_{\rm Pl}}}}\right].
\end{eqnarray}

As mentioned earlier, in order to check the model during the
inflation, one needs to obtain the number of e--folding that
somehow describes the rate of the expansion, and is defined as
\begin{equation}
N=\int_{t_{i}}^{t_{e}} Hdt=\int_{\phi_{i}}^{\phi_{e}} \frac{H}{\dot{\phi}}d\phi,
\end{equation}
where the subscripts ``i" and ``e" denote the beginning and the
end of the inflation, respectively. By using equations
(\ref{appro}) and (\ref{approo}), we can easily get
\begin{eqnarray}\label{efold}
N\!\!\!\!\!\!&\approx\!\!\!\! &\!\!\!-\frac{1}{M^{2}_{\rm
Pl}}\int_{\phi_{i}}^{\phi_{e}} \frac{V_{\rm eff}(\phi)}{V'_{\rm
eff}(\phi)}d\phi \nonumber \\ \!\!\!&=&\!\!\!-\frac{1}{M^{2}_{\rm
Pl}}\int_{\phi_{i}}^{\phi_{e}} \frac{\frac{\lambda
M^{4+n}}{\phi^{n}}+\rho
e^{\frac{\left(1-3w\right)\beta\phi}{M_{\rm Pl}}}}{-n\frac{\lambda
M^{4+n}}{\phi^{n+1}}+\frac{\left(1-3w\right)\beta}{M_{\rm Pl}}\rho
e^{\frac{\left(1-3w\right)\beta\phi}{M_{\rm Pl}}}}d\phi,
\end{eqnarray}
where the last line is obtained by substituting potential
(\ref{5}) into the effective potential and its derivative. In
addition to the number of free parameters used in this integral,
we are supposed to evaluate it for the model in hand that makes it
not to be solved easily as it stands. In other words, although
$\rho$ is independent of $\phi$, both of them are engaged through
the equations of motion.

Hence,  while using the corresponding equations of the model, we
proceed by indicating the coupling term $\rho
\,\exp{\left[(1-3w)\beta\phi/M_{\rm Pl}\right]}$ in terms of
$\phi$ alone. To perform this task, we start from relation
(\ref{ett}) and rewrite it as
\begin{equation}\label{ett11}
\ddot{\phi}+(\eta_{1}-\epsilon)H\dot{\phi}-\frac{(1+w)(1-3w)\beta\rho
e^{\left(1-3w\right) \frac{\beta\phi}{M_{\rm Pl}}}}{M_{\rm
Pl}}\approx 0.
\end{equation}
According to Ref.~\cite{brax}, the chameleon is slow rolling along
the minimum of the effective potential and hence, follows the
attractor solution $\phi\approx\phi_{\rm min}$ as long as
$m_{\phi}\gg H$. Such a condition, using equation (\ref{approo})
and definition $m^{2}_{\phi}=V''_{\rm eff}(\phi_{\rm min})$, leads
to $\eta_{1}\gg1$\rlap.$^{1}$\footnotetext[1]{As, through the
numerical analysis in finding appropriate values for a successful
inflation, the last term in relation (\ref{ett}) attains a
negative large value, hence, this condition on $\eta_{1}$
should~not, in general, prevent the slow--roll condition
$|\eta|\ll1$.}\
 Hence, by considering it into relation
(\ref{ett11}), one can neglect $\epsilon$ with respect to
$\eta_{1}$, and in turn, neglects the term $\ddot{\phi}$ with
respect to $\eta_{1}H\dot{\phi}$ term due to condition (\ref{ap}).
Therefore, relation (\ref{ett11}) reads
\begin{equation}\label{ett111}
\eta_{1}H\dot{\phi}-\frac{(1+w)(1-3w)\beta\rho
e^{\left(1-3w\right) \frac{\beta\phi}{M_{\rm Pl}}}}{M_{\rm
Pl}}\approx 0.
\end{equation}
At this stage, by substituting definition (\ref{epsilon11}) and
equations (\ref{appro}) and (\ref{approo}) into relation
(\ref{ett111}), and performing some manipulations, it yields
\begin{eqnarray}\label{10}
\frac{(1-3w)\beta}{M_{\rm Pl}}A\rho^{2}e^{2\left(1-3w\right)
\frac{\beta\phi}{M_{\rm Pl}}}\!+\!\frac{(1-3w)\beta}{M_{\rm
Pl}}B_{+}(\phi)\rho e^{\left(1-3w\right)\frac{\beta\phi}{M_{\rm
Pl}}}+V'V''\approx 0.
\end{eqnarray}
Then, by considering $w\neq 1/3$, the following two solutions
corresponding to positive and negative parts are obtained as
\begin{eqnarray}\label{rrrho}
\rho e^{\left(1-3w\right) \frac{\beta\phi}{M_{\rm Pl}}}\approx
\frac{-B_{+}(\phi)\pm\sqrt{B_{-}(\phi)^{2}+C(\phi)}}{2A},
\end{eqnarray}
where
\begin{eqnarray}\label{100}
A\!\!\!&\equiv \!\!\!&\frac{1}{M^{2}_{\rm Pl}}\left[(1-3w)^{2}\beta^{2}+3(1+w)\right],\nonumber
\\
B_{\pm}(\phi)\!\!\!&\equiv \!\!\!&V''\pm\frac{(1-3w)\beta}{M_{\rm Pl}}V'+\frac{3(1+w)}{M^{2}_{\rm Pl}}V,\nonumber
\\
C(\phi)\!\!\!&\equiv \!\!\!&\frac{12(1-3w)(1+w)\beta}{M_{\rm
Pl}}V'\left[\frac{V}{M^{2}_{\rm Pl}}-
\frac{V''}{(1-3w)^{2}\beta^{2}}\right].
\end{eqnarray}
By employing the typical potential (\ref{5}), the last two definitions can be rewritten as
\begin{eqnarray}\label{1000}
B_{\pm}(\phi)=\frac{\lambda M^{4+n}}{\phi^{n}}B^{*}_{\pm}(\phi),\nonumber
\\
C(\phi)=\frac{\lambda^2 M^{2(4+n)}}{\phi^{2n}}C^{*}(\phi)
\end{eqnarray}
and thus, relation (\ref{rrrho}) is
\begin{eqnarray}\label{rho}
\rho e^{\left(1-3w\right)\frac{\beta\phi}{M_{\rm
Pl}}}\approx\frac{\lambda
M^{4+n}}{\phi^{n}}F^{(\pm)}(\phi)=V(\phi)F^{(\pm)}(\phi),
\end{eqnarray}
where $F^{(\pm)}$ correspond to positive and negative parts of the solutions, and are defined as
\begin{eqnarray} \label{f}
F^{(\pm)}(\phi)\equiv\frac{-B^{*}_{+}(\phi)\pm\sqrt{B^{*\,
2}_{-}(\phi)+C^{*}(\phi)}}{2A},
\end{eqnarray}
in which
\begin{eqnarray}\label{1000}
B^{*}_{\pm}(\phi)\!\!\!&\equiv \!\!\!&\frac{n(n+1)}{\phi^{2}}\mp
n\frac{(1-3w)\beta}{M_{\rm Pl}\phi}+ \frac{3(1+w)}{M^{2}_{\rm
Pl}},\nonumber
\\
C^{*}(\phi)\!\!\!&\equiv \!\!\!&-\frac{12n(1-3w)(1+w)\beta}{M_{\rm
Pl}}\left[\frac{1}{{M^{2}_{\rm Pl}}\phi}-
\frac{n(n+1)}{(1-3w)^{2}\beta^{2}\phi^{3}}\right].
\end{eqnarray}
In fact, through the chameleon condition and some plausible
approximations, relation (\ref{rho}) indicates that the coupling
term $\rho \,\exp{\left[(1-3w)\beta\phi/M_{\rm Pl}\right]}$ has
been taken to be proportional to $V(\phi)$, which leads to somehow
relating the matter density as a function of the scalar field.
However, on the other point of view, we are actually proceeding
the work, as if one just continues the work by imposing relation
(\ref{rho}) as {\it a priori} assumption. Moreover, using this
relation results in the elimination of the free parameters $M$ and
$\lambda$ in the subsequent calculations.

Now, let us substitute relations (\ref{rho}) into (\ref{ep}) and (\ref{efold}) to get
\begin{eqnarray}\label{epf}
\epsilon^{(\pm)}(\phi)\approx \frac{M^{2}_{\rm
Pl}}{2}\left[\frac{\frac{-n}{\phi}+ \frac{(1-3w)\beta}{M_{\rm
Pl}}F^{(\pm)}(\phi)}{1+F^{(\pm)}(\phi)}\right]^{2}+\frac{3}{2}\left[
\frac{(1+w)F^{(\pm)}(\phi)}{1+F^{(\pm)}(\phi)}\right]
\end{eqnarray}
and
\begin{eqnarray}\label{efold1}
N^{(\pm)}(\phi)\approx -\frac{1}{M^{2}_{\rm
Pl}}\int_{\phi_{i}}^{\phi_{e}}\frac{1+F^{(\pm)}(\phi)}{\frac{-n}{\phi}+\frac{(1-3w)\beta}{M_{\rm
Pl}}F^{(\pm)}(\phi)}d\phi,
\end{eqnarray}
where $\epsilon^{(\pm)}(\phi)$ and $N^{(\pm)}(\phi)$ correspond to
two different values of $F^{(\pm)}(\phi)$ defined in (\ref{f}).
Furthermore, in order to get the value of $\phi$ at the end of the
inflation, we use the known relation
\begin{equation}\label{e1}
\epsilon|_{\phi=\phi_{e}}\approx 1,
\end{equation}
that means the inflation ends when the slow--roll scenario breaks
down by growing $\epsilon$ up to the order of one. Moreover, for
indicating the value of $\phi_{i}$ at the beginning of inflation,
we employ$^{1}$\footnotetext[1]{The consistency relation
(\ref{i1}) is usually used for a single--field model, however,
since we are~not building a fully realistic model, as an
approximation, we have restricted ourselves to it and also to its
observational constraint for reason of simplicity in getting a
rough estimation of $\phi_{i}$. In fact, as mentioned below
relation (\ref{1000}), employing the consistency relation
(\ref{i1}) would be plausible. Nevertheless, and generally
speaking, it may affect the analysis, although, by the argument
mentioned at the last paragraph of Sect.~$3$, one expects that it
would play a minor role in the results. Moreover, since in a
two--field model, the value of $r$ is less than its value in a
single--field one~\cite{wand,wand1,wand2}, it does not ruin the
results of the work as explained in Footnote $1$ on Page
$12$.}~\cite{liddle,ruth}\
 the relation
\begin{equation}\label{i1}
r\approx 16\epsilon|_{\phi=\phi_{i}},
\end{equation}
where the parameter $r$ is the ratio of the tensor perturbation
amplitude to the scalar perturbation amplitude. The Planck
temperature anisotropy measurements have released an upper limit
for this parameter to be $r<0.11$ in $95\%$ confidence
level~\cite{p13}--\cite{p152}.

Then, by using relations (\ref{epf}), (\ref{e1}) and (\ref{i1}),
one can obtain the values of $\phi$ at the beginning and at the
end of inflation, therefore the number of e--folding can be
estimated numerically solving integral
(\ref{efold1})\rlap.$^{2}$\footnotetext[2]{In order to scan the
entire phase--space of solutions, the both values of $N^{(+)}$ and
$N^{(-)}$ have been used in the analysis.}\
 Through this analysis,
we not only investigate the viability of the model by indicating
the admissible number of e--folding, but also, we attempt to set
constraints to pin down the free parameters of the model. As
mentioned, the integral $N$ consists of many free parameters and
hence, not easy to be solved analytically. In this regard, in what
follows, through some values of $\beta$, $n$, $w$ and $r$, we
estimate the integral numerically to attain admissible values for
the number of e--folding. We work in an appropriate unit to set
$M_{\rm Pl}=1$. Also, for choosing plausible values of $\beta$, we
have noted that Weltman and Khoury, in their original suggestion
for the chameleon model~\cite{weltman1,weltman2}, considered the
possibility of coupling the scalar field to the matter field with
the gravitational strength, and have shown that the chameleon
theory is compatible while the coupling constant, $\beta$, is of
the order of unity. However, Mota and Shaw have shown that the
scalar field theories, which couple to the matter field much more
strongly than gravity, are viable due to the non--linearity
effects of the theory~\cite{mota1,mota2}. Thus, we assume
$\beta=0.1$, $1$ and $1000$ to cover more possible coupling
strength. In addition, by taking $r=0.109$, $-1\leqslant
w\leqslant 1$ and different values of $n$, we estimate the number
of e--folding for these values of $\beta$. In this respect, and in
the first step, we have depicted the behavior of e--folding $N$
with respect to $n$ and $w$ in Figs.~$1$ and $2$. For example, for
$\beta=1$, we illustrate the behavior of $N^{(-)}$ with respect to
$n$ for some fixed values of $w$ with few plots in Fig.~$1$, and
also with respect to $w$ for some fixed values of $n$ with few
plots in Fig.~$2$. The results indicate that the power--law
potentials with $n<0$, as a chameleon self--interacting potential
(except $n=-2$ and odd--negative integers), with the equation of
state parameters near $-1$ for the coupled matter field seem to be
more compatible with the inflation. In Fig.~$2$, the singularities
around $w\approx 1/3$ are due to the fact that the chameleon field
does~not couple to the radiation. Also note that, the plots in
Fig.~$1$ have been drawn for the continuous range of parameter $n$
whereas we have assumed $n$ to be integer for the typical
potential of the model. In order to attain an admissible number of
e--folding more precisely, we calculate the two solutions
$N^{(\pm)}$ for different chosen values of the free parameters
$n$, $w$ and $\beta$. The results have been collected in
Tables~$1$--$5$, which indicate that for having a viable
inflationary model, the self--interacting potential in the form of
power--law with $n=-2$ and the equation of state parameter of the
coupled field with $w=-1$ seems to be more appropriate than the
other choices. However, the case $n=-2$ is~not suitable as a
chameleonic potential because it is linear and does~not enable the
model to be screened~\cite{t8,Burrage2017}. Note that, the range
of allowable $w$ tends to be larger for strongly coupling case
(i.e., $\beta=1000$) than the other cases.
\begin{figure}[htbp]
\includegraphics[scale=0.4]{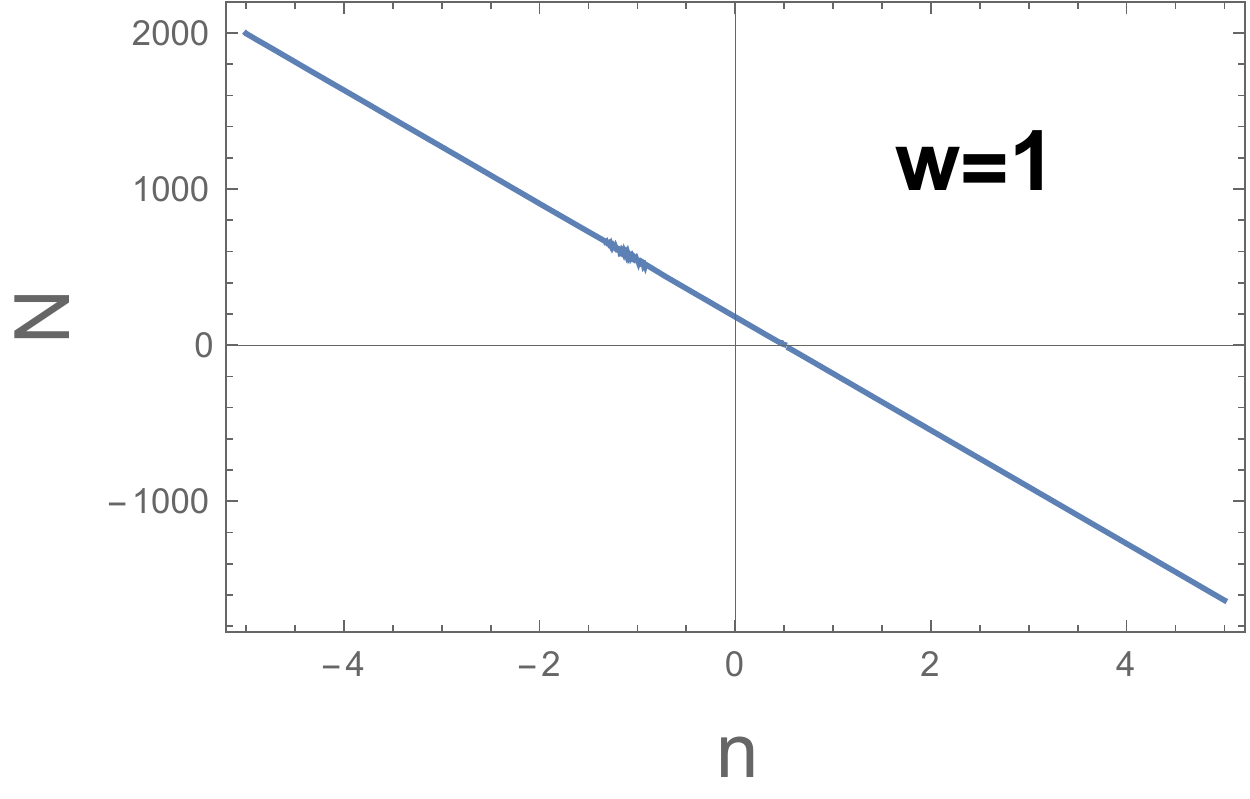}
\includegraphics[scale=0.4]{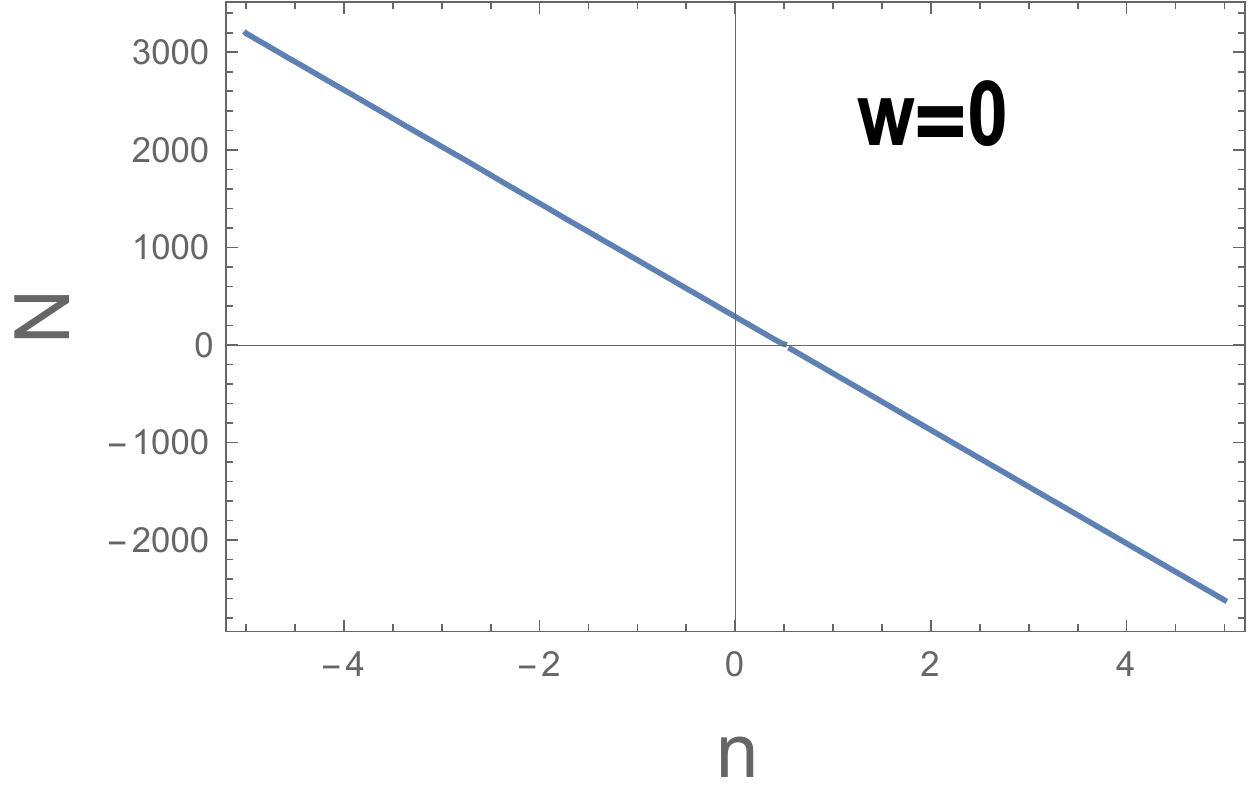}
\includegraphics[scale=0.4]{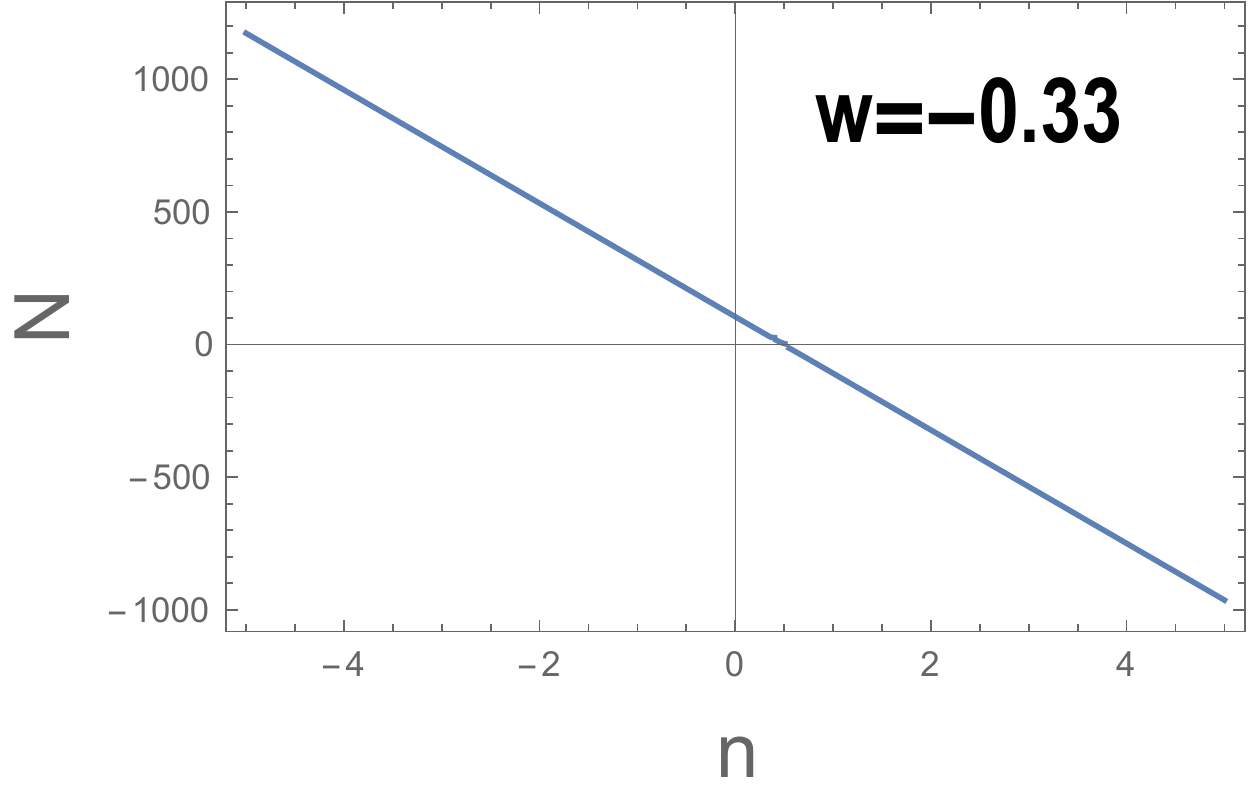}
\includegraphics[scale=0.4]{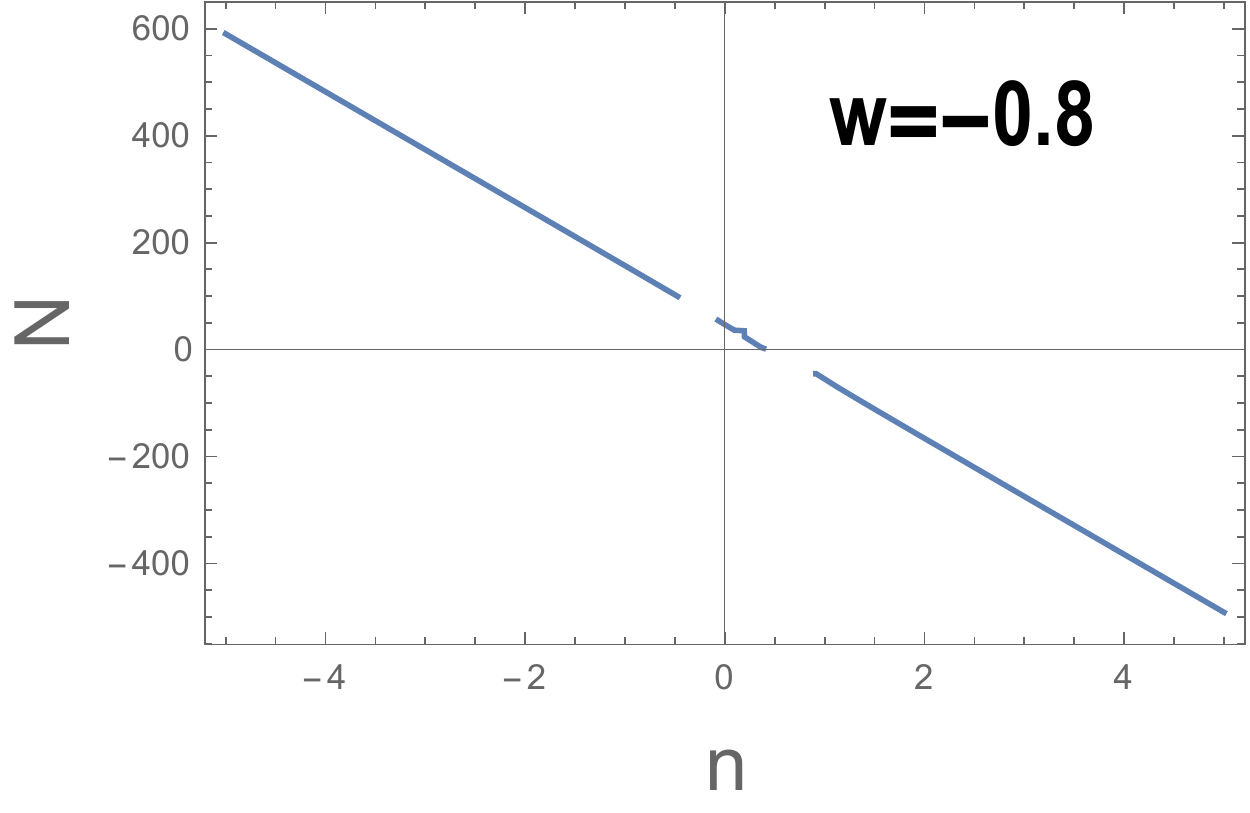}
\includegraphics[scale=0.4]{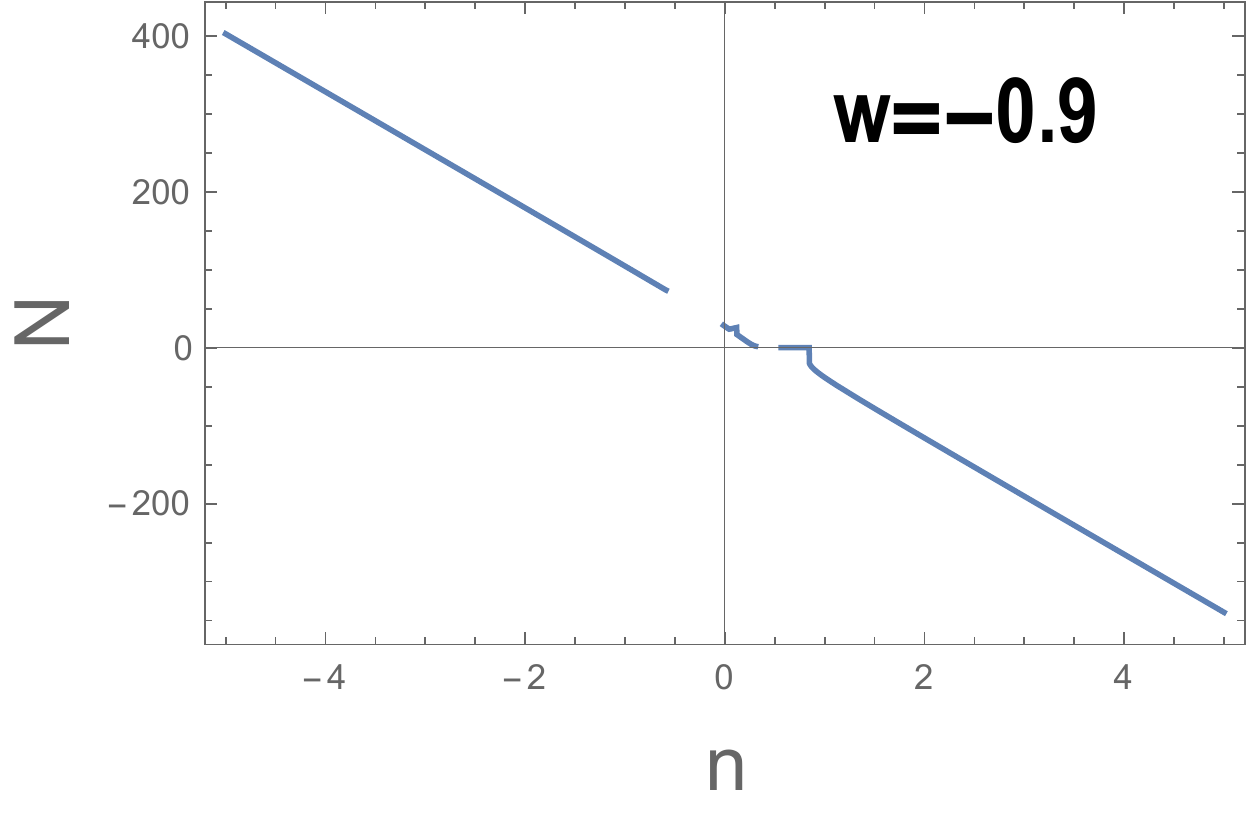}
\includegraphics[scale=0.4]{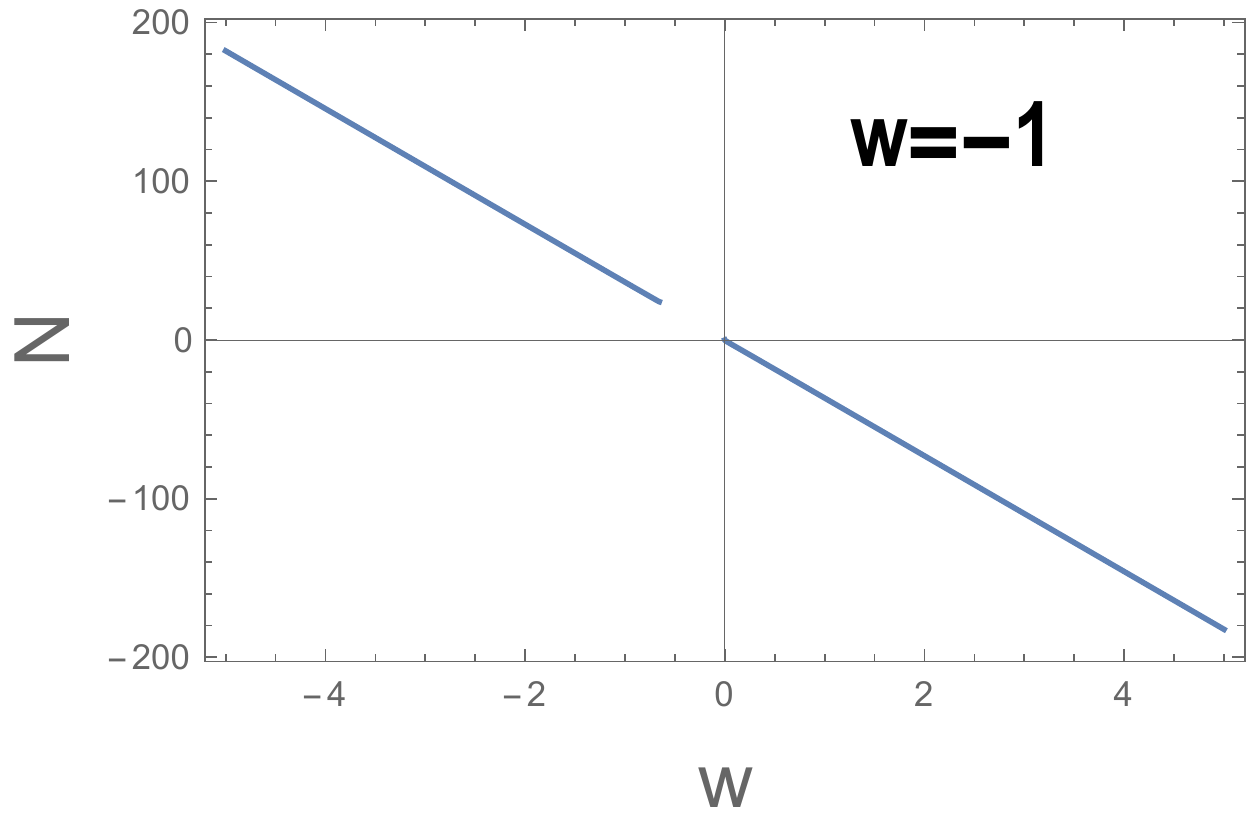}
\caption{\label{}\small The figures show the behavior of
e--folding $N^{(-)}$ with respect to $n$ for different values of
$w$. As seen, the power--law potentials with $n<0$, as a
self--interacting potential, with the equation of state parameters
near $-1$ for the coupled matter field seem to be more suitable
during the inflation. Also, we have set $\beta=1$, $r=0.109$ and
$M_{\rm Pl}=1$ in an appropriate unit. }
\end{figure}
\begin{figure}[htbp]
\includegraphics[scale=0.4]{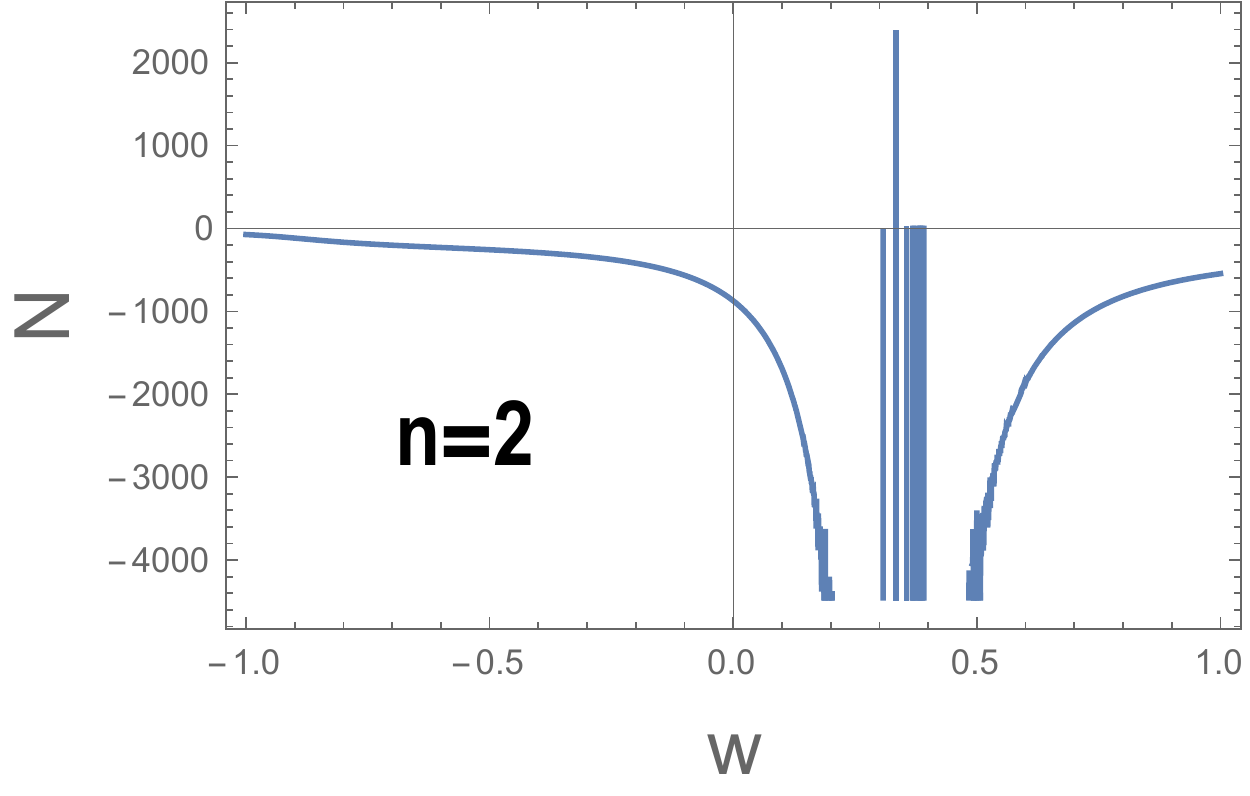}
\includegraphics[scale=0.4]{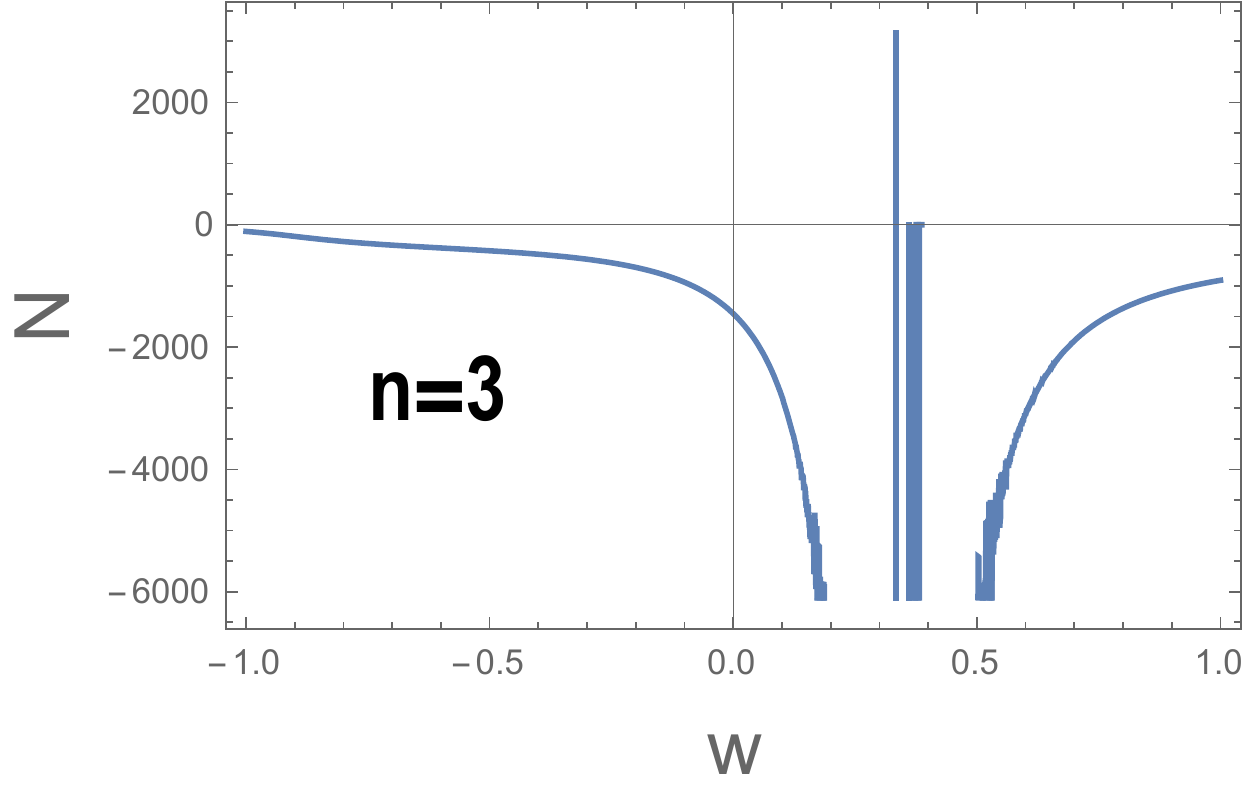}
\includegraphics[scale=0.4]{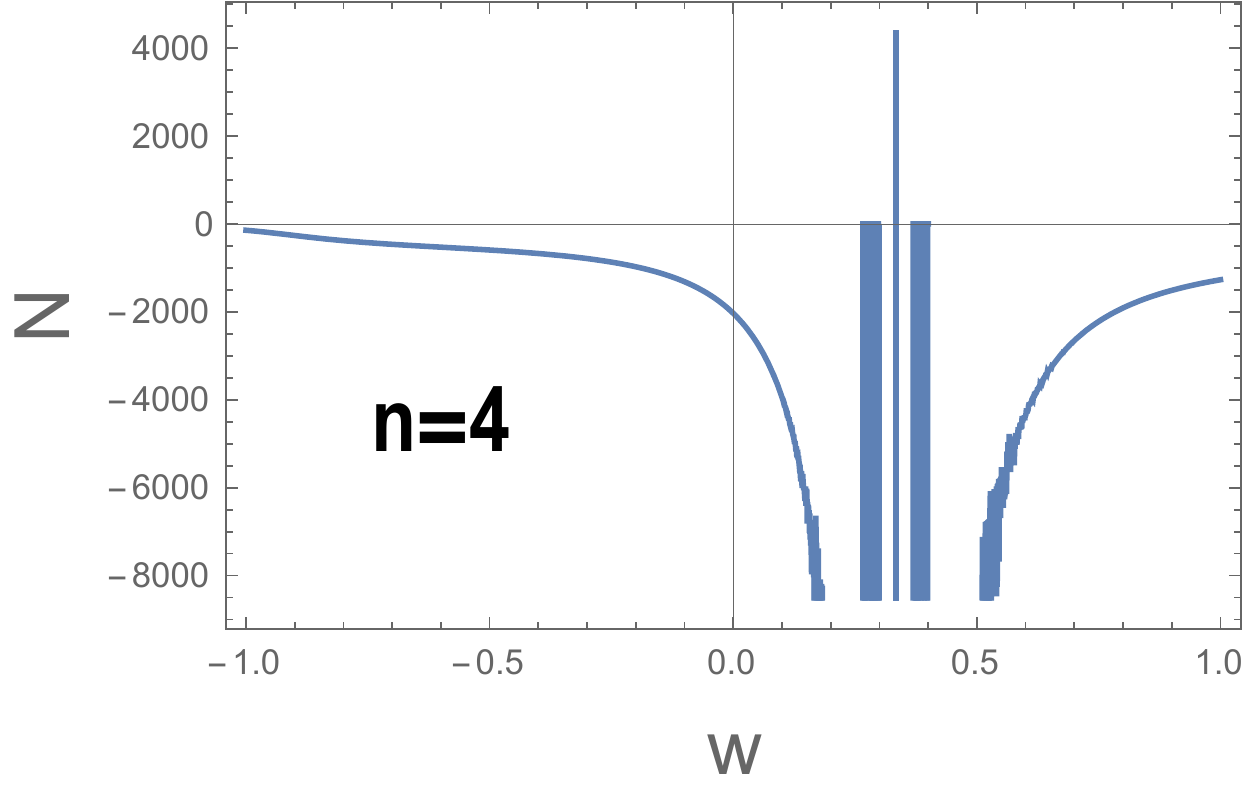}
\includegraphics[scale=0.4]{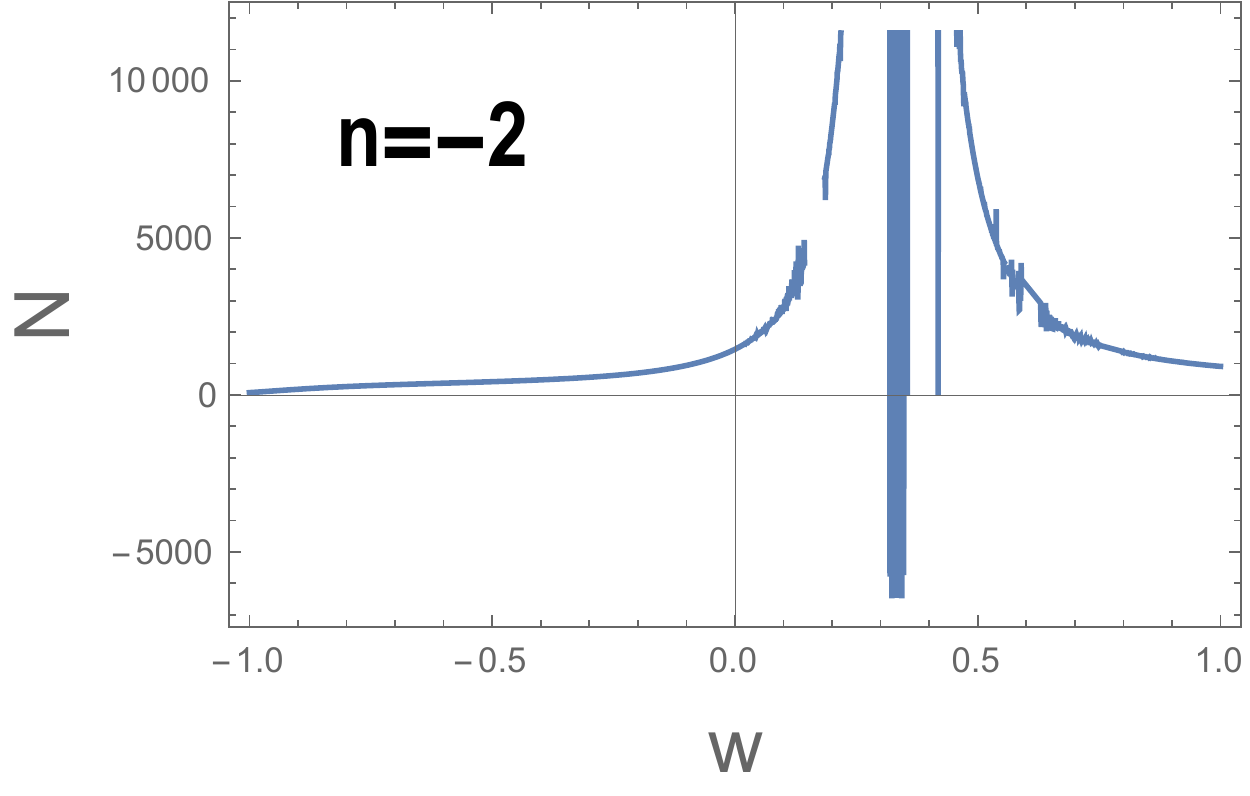}
\includegraphics[scale=0.4]{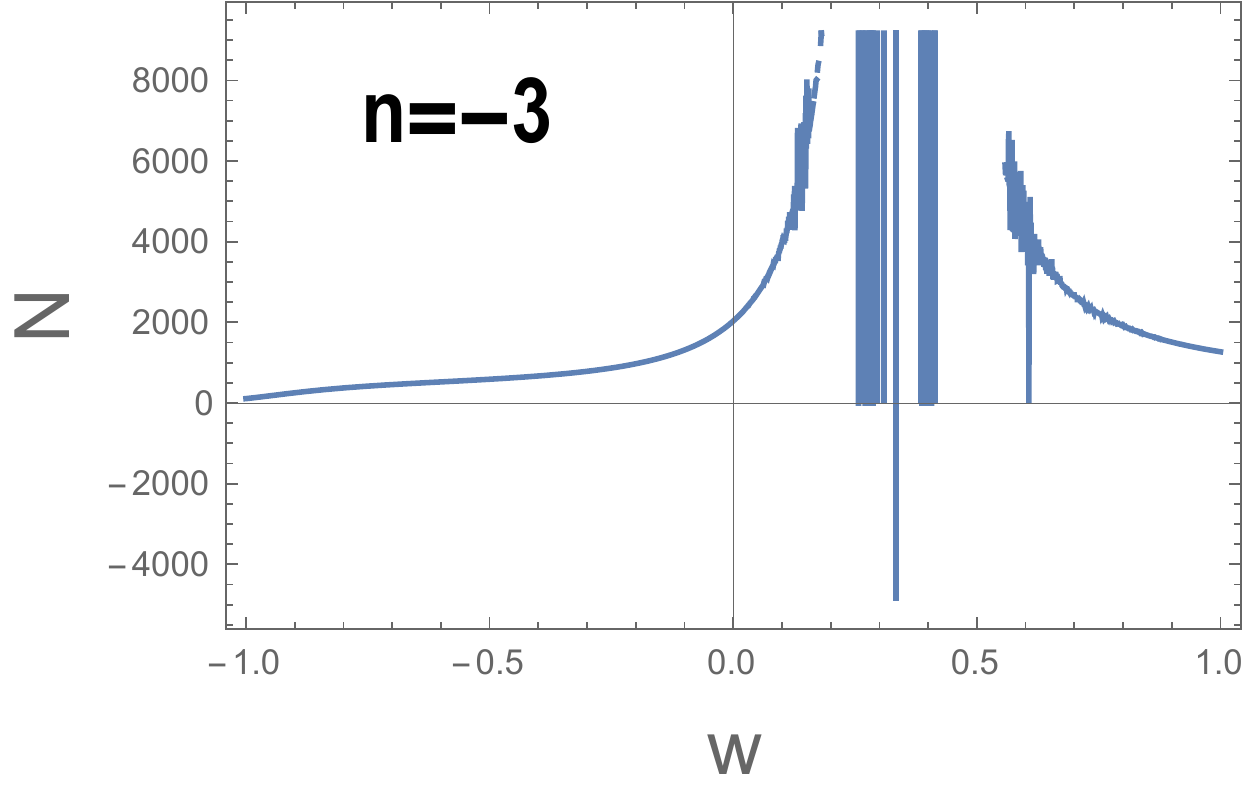}
\includegraphics[scale=0.4]{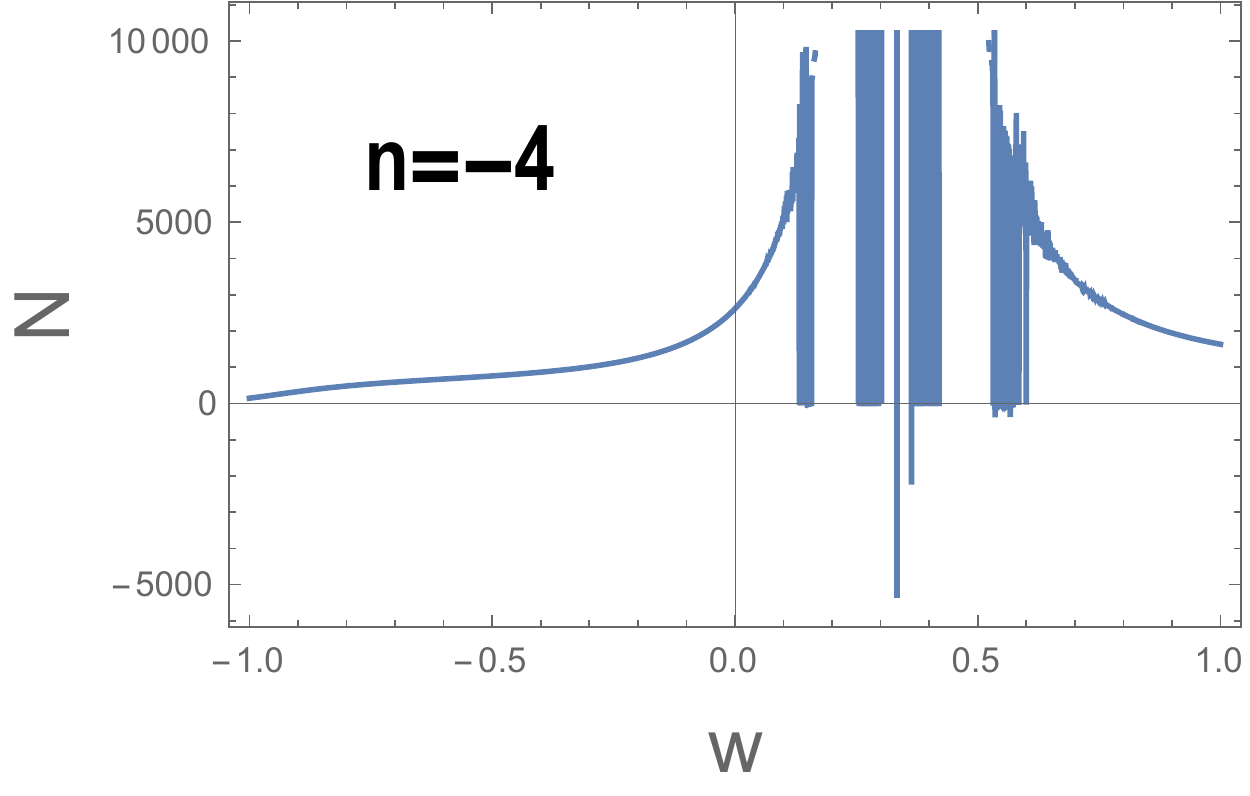}
\caption{\label{}\small The figures show the behavior of
e--folding $N^{(-)}$ with respect to the equation of state
parameter of the coupled matter field $w$ for some different
values of $n$. As seen, the power--law potentials with $n<0$, as a
self--interacting potential, with the equation of state parameters
near $-1$ for the coupled matter field seem to be more suitable
during the inflation. Since the chameleon field does~not directly
couple to the radiation, the plots show some singularities near
$w\approx1/3$. Also, we have set $\beta=1$, $r=0.109$ and $M_{\rm
Pl}=1$ in an appropriate unit. }
\end{figure}

Furthermore, there is another constraint in the chameleon model,
i.e., $\beta V'(\phi_{\rm min})<0$ as mentioned, e.g., in
Refs.~\cite{weltman1,mota2}. It has been obtained due to $V'_{\rm
eff}(\phi_{\rm min})\approx 0$ that leads to
\begin{equation}\label{phii}
\phi_{\rm min}\approx\left[\frac{n\lambda M^{4+n}M_{\rm
Pl}}{(1-3w)\beta\rho e^{(1-3w)\frac{\beta\phi_{\rm min}}{M_{\rm
Pl}}}}\right]^{\frac{1}{n+1}}.
\end{equation}\\
This relation implies that, if $\beta <0$, then $n$ can be either
a negative integer or an even--positive integer. And, if
$\beta>0$, then $n$ can be either a positive integer or an
even--negative integer. Thus, without lose of generality, as it is
common in the chameleon models, we only restrict the analysis on
the cases with $\beta>0$, that it is also consistent with the
allowed integers $n$ mentioned in, e.g.,
Refs.~\cite{t8,Burrage2017}.
\begin{table}[htbp]
  \centering
  {\footnotesize
  \begin{tabular}{||c||c|c|c|c|c|c|c||}
\hline \hline
    \multirow{2}{*}{$\bf {n}$} & \multicolumn{7}{c||}{${\bf N^{(-)}}$}        \\ \cline{2-8}
 &${\bf w=-1}$&${\bf w=-0.9}$ &${\bf w=-0.8}$ &${\bf w=-0.7}$&${\bf w=-0.3}$&${\bf w=0}$&${\bf w=1}$\\\hline \hline
   ${\bf -4}$ &     $143.668$  & $2003.07$ &$4015.54$ &$6740.1$&$ $&$ $&$2.93124\times10^{-8}$   \\ \hline
     ${\bf -2}$ &     ${\bf 70.9938}$  & $1111.48$ &$2229.54$ &$3793.26$&$ $&$109701$&$ $   \\ \hline
       ${\bf 1}$ &     $-38.8159$  & $-225.92$ &$-449.104$ &$-754.772$&$-4313.4$&$-21041.4$&$-11375.4$   \\ \hline
        ${\bf 2}$ &     $-75.2381$  & $-670.98$ &$-1341.94$ &$-2259.84$&$ $&$1.0645\times10^{-6}$&$-33020.3$   \\ \hline
         ${\bf 3}$ &    $-111.663$  & $-1116.66$ &$-2234.95$ &$-3764.94$&$-21558.7$&$ $&$-70474.9$   \\ \hline
          ${\bf 4}$ &     $-148.096$  & $-1562.37$ &$-3127.95$ &$-5270.71$&$ $&$-139306$&$2.29345\times10^{-8}$   \\ \hline
           ${\bf 5}$ &     $-184.532$  & $-2008.09$ &$-4020.94$ &$-6773.86$&$ $&$ $&$9.14092\times10^{-9}$   \\ \hline\hline

  \end{tabular}
  }
  \caption{The Table shows the number of e--folding
$N^{(-)}$ for different values of $n$ and $w$. The values
corresponding to the blank cells are imaginary and have been
ignored. It seems that the self--interacting potential with
$n=-2$, and with the equation of state parameter $w=-1$ (related
to the coupled matter field), is about to be an appropriate value,
however, it is~not suitable as a chameleonic potential. Also, we
have set ${\bf\beta=0.1}$, $r=0.109$ and $M_{\rm Pl}=1$ in an
appropriate unit. Note that, we have~not found any solutions for
$N^{(+)}$ in this case.}\label{Tab03}
\end{table}
\begin{table}[htbp]
  \centering
  {\footnotesize
  \begin{tabular}{||c||c|c|c|c|c|c|c||}
\hline \hline
   \multirow{2}{*}{$\bf {n}$}  & \multicolumn{7}{c||} {${\bf N^{(-)}}$}        \\ \cline{2-8}
 &${\bf w=-1}$&${\bf w=-0.9}$ &${\bf w=-0.8}$ &${\bf w=-0.7}$&${\bf w=-0.3}$&${\bf w=0}$&${\bf w=1}$\\ \hline\hline
   ${\bf -4}$ &     $145.705$  & $328.576$ &$482.15$ &$588.954$&$1011.36$&$2612.69$&$1633.45$   \\ \hline
     ${\bf -2}$ &     ${\bf 72.833}$  & $179.678$ &$265.481$ &$325.145$&$561.165$&$1452.17$&$907.681$   \\ \hline
       ${\bf 1}$ &     $-36.6075$  & $-38.0129$ &$-55.3592$ &$-69.602$&$-113.988$&$-291.337$&$-182.108$   \\ \hline
        ${\bf 2}$ &     $-73.0298$  & $-115.665$ &$-166.071$ &$-201.83$&$-339.089$&$-872.33$&$-545.23$   \\ \hline
         ${\bf 3}$ &     $-109.467$  & $-190.469$ &$-274.668$ &$-333.801$&$-564.19$&$-1453.26$&$-908.318$   \\ \hline
          ${\bf 4}$ &     $-145.909$  & $-264.978$ &$-383.048$ &$-465.711$&$-789.284$&$-2034.27$&$-1271.44$   \\ \hline
           ${\bf 5}$ &     $-182.353$  & $-339.394$ &$-491.36$ &$-597.6$&$-1014.37$&$-2615.15$&$-1634.56$   \\ \hline\hline

  \end{tabular}
  }
  \caption{The Table shows the number of e--folding $N^{(-)}$ for different values of
$n$ and $w$. It seems that the self--interacting potential with
$n=-2$, and with the equation of state parameter $w=-1$ (related
to the coupled matter field), is about to be an appropriate value,
however, it is~not suitable as a chameleonic potential. Also, we
have set ${\bf\beta=1}$, $r=0.109$ and $M_{\rm Pl}=1$ in an
appropriate unit.}\label{Tab01}
\end{table}
\begin{table}[htbp]
  \centering
  {\footnotesize
  \begin{tabular}{||c||c|c|c|c|c|c|c||}
\hline \hline
   \multirow{2}{*}{$\bf {n}$}  & \multicolumn{7}{c||} {${\bf N^{(+)}}$}        \\ \cline{2-8}
 &${\bf w=-1}$&${\bf w=-0.9}$ &${\bf w=-0.8}$ &${\bf w=-0.7}$&${\bf w=-0.3}$&${\bf w=0}$&${\bf w=1}$\\ \hline\hline
   ${\bf -4}$ &     $145.599$  & $190.408$ &$-6008.3$ &$ $&$38017$&$13920.5$&$48476.1$   \\ \hline
     ${\bf -2}$ &     ${\bf 72.7919}$  & $-1378.57$ &$-1274.6$ &$-3046.55$&$95265.8$&$10925.8$&$43415.6$   \\ \hline
       ${\bf 1}$ &     $ $& $ $ &$ $ &$ $&$ $&$ $
       &$-41.6261$   \\ \hline
        ${\bf 2}$ &     $ $  & $ $ &$ $ &$ $&$-1120.39 $&$ $&$-2565.36$   \\ \hline
         ${\bf 3}$ &     $ $  & $ $ &$ $ &$ $&$-3261.07$&$-2629.54$&$-7250.38$   \\ \hline
          ${\bf 4}$ &     $ $  & $ $ &$ $ &$ $&$-5917.94$&$-4458.47$&$-12717.2$   \\ \hline
           ${\bf 5}$ &     $ $  & $ $ &$ $ &$ $&$-8858.71$&$-6379.66$&$-18551.6$   \\ \hline\hline

  \end{tabular}
  }
  \caption{The Table shows the number of e--folding $N^{(+)}$ for different values
of $n$ and $w$. The values corresponding to the blank cells are
imaginary and have been ignored. It seems that the
self--interacting potential with $n=-2$, and with the equation of
state parameter $w=-1$ (related to the coupled matter field), is
about to be an appropriate value, however, it is~not suitable as a
chameleonic potential. Also, we have set ${\bf\beta=1}$, $r=0.109$
and $M_{\rm Pl}=1$ in an appropriate unit.}\label{Tab02}
\end{table}
\begin{table}[htbp]
  \centering
  {\footnotesize
  \begin{tabular}{||c||c|c|c|c|c|c|c||}
\hline \hline
   \multirow{2}{*}{$\bf {n}$}  & \multicolumn{7}{c||} {${\bf N^{(-)}}$}        \\ \cline{2-8}
 &${\bf w=-1}$&${\bf w=-0.9}$ &${\bf w=-0.8}$ &${\bf w=-0.7}$&${\bf w=-0.3}$&${\bf w=0}$&${\bf w=1}$\\ \hline\hline
   ${\bf -4}$ &     $145.789$  & $145.959$ &$146.16$ &$146.398$&$148.113$&$152.132$&$152.132$   \\ \hline
     ${\bf -2}$ &     ${\bf72.8945}$  & ${\bf72.9964}$ &${\bf73.1164}$ &${\bf73.2597}$&${\bf74.2885}$&$76.6999$&$76.6999$   \\ \hline
       ${\bf 1}$ &     $-36.4472$  & $-36.4472$ &$-36.4472$ &$-36.4472$&$-36.4474$&$-36.4473$&$-36.4472$   \\ \hline
        ${\bf 2}$ &     $-72.8945$  & $-72.9285$ &$-72.9684$ &$-73.0162$&$-73.3587$&$-74.1634$&$-74.1635$   \\ \hline
         ${\bf 3}$ &    $-109.342$  & $-109.41$ &$-109.49$ &$-109.585$&$-110.269$&$-111.88$&$-111.879$   \\ \hline
          ${\bf 4}$ &     $-145.789$  & $-145.891$ &$-146.011$ &$-146.154$&$-147.179$&$-149.596$&$-149.595$   \\ \hline
           ${\bf 5}$ &     $-182.236$  & $-182.372$ &$-182.532$ &$-182.729$&$-184.095$&$-187.311$&$-187.313$   \\ \hline\hline

  \end{tabular}
  }
  \caption{The Table shows the number of e--folding $N^{(-)}$ for different
values of $n$ and $w$. The values corresponding to the blank cells
are imaginary and have been ignored. It seems that the
self--interacting potential with $n=-2$, and with the equation of
state parameters $w<0$ (related to the coupled matter field), is
about to be an appropriate value, however, it is~not suitable as a
chameleonic potential. Also, we have set ${\bf\beta=1000}$,
$r=0.109$ and $M_{\rm Pl}=1$ in an appropriate unit. The results
show that the range of allowable $w$ tends to be larger for the
strongly coupling case than the other cases.}\label{Tab03}
\end{table}
\begin{table}[htbp]
  \centering
  {\footnotesize
  \begin{tabular}{||c||c|c|c|c|c|c|c||}
\hline \hline
   \multirow{2}{*}{$\bf {n}$}  & \multicolumn{7}{c||} {${\bf N^{(+)}}$}        \\ \cline{2-8}
 &${\bf w=-1}$&${\bf w=-0.9}$ &${\bf w=-0.8}$ &${\bf w=-0.7}$&${\bf w=-0.3}$&${\bf w=0}$&${\bf w=1}$\\ \hline\hline
   ${\bf -4}$ &     $145.789$  & $145.619$ &$145.419$ &$145.181$&$143.482$&$139.567$&$139.567$   \\ \hline
     ${\bf -2}$ &     ${\bf 72.8945}$  & ${\bf72.7926}$ &${\bf72.6728}$ &${\bf72.53}$&${\bf71.5104}$&${\bf69.1622}$&${\bf69.162}$ \\ \hline
       ${\bf 1}$ &     $-36.4472$  & $-36.4472$ &$-36.4472$ &$-36.4472$&$-36.4474$&$-36.4473$&$-36.4472$   \\ \hline
        ${\bf 2}$ &     $-72.8945$  & $-72.8605$ &$-72.8207$ &$-72.773$&$-72.4326$&$-71.6496$&$-71.6497$   \\ \hline
         ${\bf 3}$ &    $-109.342$  & $-109.274$ &$-109.194$ &$-109.099$&$-108.417$&$-106.852$&$-106.852$   \\ \hline
          ${\bf 4}$ &     $-145.789$  & $-145.687$ &$-145.567$ &$-145.424$&$-144.401$&$-142.055$&$-142.055$   \\ \hline
           ${\bf 5}$ &     $-182.236$  & $-182.1$ &$-181.941$ &$-181.756$&$-180.391$&$-177.257$&$-177.259$   \\ \hline\hline

  \end{tabular}
  }
  \caption{The Table shows the number of e--folding $N^{(+)}$ for different
values of $n$ and $w$. The values corresponding to the blank cells
are imaginary and have been ignored. It seems that the
self--interacting potential with $n=-2$, and with the equation of
state parameter $w=-1$ (related to the coupled matter field), is
about to be an appropriate value, however, it is~not suitable as a
chameleonic potential. Also, we have set ${\bf\beta=1000}$,
$r=0.109$ and $M_{\rm Pl}=1$ in an appropriate unit. The results
show that the range of allowable $w$ tends to be larger for the
strongly coupling case than the other cases.}\label{Tab03}
\end{table}

Meanwhile, we have probed the coupled matter field case for $n=-2$
(although not as a chameleon model) to find its corresponding
energy density through our imposed relation (\ref{rho}). In this
regard, the results of calculations indicate negative values for
its energy density, where such possibility may be related to some
exotic matters which are hypothetical forms of the matters for
describing the wormholes~\cite{vi}. Although in the recent years,
some efforts have been devoted to this topic through extensions of
the quantum field theories and also via studies of the Casimir
effects~\cite{ep}--\cite{ne}, it is~not mostly acceptable due to
the violation of various energy conditions in general relativity.
Hence, this admissible value of $n$ is also excluded in our
approach. There are also some other points to be noted, first in
Fig.~$1$, the continuous lines (in particular, for $w=0$ and
$w=-0.33$) show that the non--integers $n$ may lead to a few
allowable e--folding numbers that intrigue one to consider these
values as well. However, we noticed that those non--integers $n$
also give imaginary or negative energy densities that are~not
acceptable. Second, the numerical analysis does~not even give any
admissible value of $n$ for e--folding more than $70$. Hence, in
order to avoid lengthy tables, we have just shown the results for
some values of $n$ in the range $-5 < n\leq 5$ (as an almost
general cases of the potential). Third, the parameter $r$ has been
taken very close to its upper bound value because through some
calculations, we have found that only the results of such choice
are closer to the appropriate values of the
e--folding\rlap.$^{1}$\footnotetext[1]{Moreover, these
calculations indicate that even considering the model as a
two--field one, wherein the value of $r$ is less than its value in
a single--field model~\cite{wand,wand1,wand2}, it does~not affect
the results of the work.}

Therefore, the results of the present analysis indicate that there
is~not much chance of having the chameleonic inflation. This null
result, that has been obtained through the mathematical
considerations, may~not be far from the expectation for the
extreme case when the matter density approaches a constant value
with $w=-1$. That is, the exponential term (entered due to the
chameleonic coupling) in the effective potential (\ref{eff})
cannot inflate unless one fixes the value of the scalar, which in
turn means that the matter density must be constant (i.e., a
cosmological constant), namely it itself can start the inflation
at the beginning without the effect of a coupling being
efficient\rlap.$^{2}$\footnotetext[2]{Meanwhile, this argument
indicates that it had been plausible to employ the consistency
relation (\ref{i1}) and its corresponding observational constraint
at the beginning of inflation.}

Nevertheless, in this work, we have considered a generalization of
the usual case. In fact, even in the limiting case where
$\beta\rightarrow 0$ (i.e., reducing to the minimal coupling
case), as the $V'_{\rm eff}(\phi)$ term in equation (\ref{phi})
reads $V'_{\rm eff}(\phi)\approx
V'(\phi)+\left[(1-3w)\frac{\beta}{M_{\rm Pl}}\right]\rho $, it
looks as if this unknown fluid somehow acts like an extra field
(where such kind of effects have been studied in, e.g.,
Ref.~\cite{wand2015}), and hence, it would be worth to explicitly
investigate any possible effect of the non--minimal coupling term
in the analysis of the extreme case.

\section{Conclusions}
\indent

In this work, we have addressed the analysis of the role of
chameleon field and its influence on the dynamics of the universe
inflation. The chameleon field, a scalar canonical field, was
introduced in the Einstein gravitational theory for the solution
of the problem of the EP--violation. The interaction of chameleon
field with an ambient matter goes through the conformal factor
that leads to a dependence of the chameleon field mass on the
matter density. We have focused on the possibility of the
chameleon field influence during the period of inflation in the
very early universe. To perform such a task, we have considered a
coupling between the chameleon scalar field and an unknown matter
scalar field with the equation of state parameter $w$ during the
inflation, wherein we have employed the common typical potentials
usually used for the chameleon gravity in the literature. In the
context of the slow--roll approximations, we obtained the
slow--roll conditions for the model that are different from the
corresponding ones of the standard model due to the non--minimal
coupling term. In order to check the ability of resolving the
problems of the standard big bang cosmology such as the flatness,
horizon and monopole, we got the number of e--folding for the
model. The calculations led us to an integral for the e--folding
that has~not been easy to be solved analytically due to the
engaged equations and lots of the free parameters. Thus, by some
plausible approximations and through some analysis and
manipulations, we managed the coupling term in the effective
potential being proportional to the potential, however, this could
also be assumed as an assumption. Hence, we somehow reduced the
free parameters to be just $n$ (the slope of the potential), $w$
(the equation of state parameter of the coupled matter field) and
$\beta$ (the coupling constant between the chameleon field and the
matter field). However, the price of this resolution is somehow
affected on not specifying the unknown matter fluid.

Then, using the observable parameter $r$, we obtained the number
of e--folding numerically for some different values of the
remained free parameters\rlap.$^{3}$\footnotetext[3]{Only two
observable parameters have been considered in this work, namely
the number of e--folding and the parameter $r$. There are also a
few other observable parameters from inflation (such as the
amplitudes of scalar perturbations evaluated while leaving the
horizon and the spectral indices) that are important in the
inflationary investigation. However, the investigated chameleon
inflationary model has~not shown much success in this approach,
hence we did~not continue to consider the other parameters in this
work.}\
 At the first step of
the analysis, we found some values of $n$ and $w$ that give a
closer value to the acceptable e--folding number, however, these
are~not accepted as a chameleon model. Besides, such values
resulted in the negative energy densities, that are~not also
acceptable due to the violation of various energy conditions in
general relativity. Hence, the remaining best values, that could
lead to successful inflationary models, have been excluded too.
Therefore, through the general form of the common typical
potential (that usually used in the context of the chameleon model
in the literature), we have provided a critical analysis that
shows there may be not much room for having a viable chameleonic
inflationary model. This conclusion overlaps with the results
obtained in Refs.~\cite{wang2,khoury}. However, still encouraged
by the results of Refs.~\cite{ivanov,ivanov2}, and the argument
presented just before the conclusions, we have been investigating
a possible approach to realize even the insignificant influence of
the chameleon field on the universe inflation. In this respect, to
retain the chameleon mechanism during the inflation, we have
suggested the following scenario. Knowing the fact that, if
through some mechanism, the chameleon inflationary model (that
consists of both the non--minimal and the minimal coupling terms)
reduces to the standard inflationary model without the
non--minimal coupling term during the inflation, then it can
obviously cover the inflationary epoch. Hence, in this regard in
Ref.~\cite{saba}, by appealing to the noncommutativity as that
mechanism, we have shown that there is a correspondence between
the chameleon model and the noncommutative standard model in the
presence of a particular type of dynamical deformation between the
canonical momenta of the scale factor and of the scalar field, and
at last, we have reached to the point that during the inflation,
the chameleon field acts~\cite{saba}. Also, as the price of the
present work resulted in somehow not specifying the unknown matter
field, this issue has been investigated in more detail in
Ref.~\cite{saba} wherein the matter field is obtained to be as a
cosmic string fluid. Nevertheless, the investigation of the
chameleon model when the universe reheats after inflation can also
be of much interest that we propose to study in a subsequent work.
Furthermore, in another work~\cite{RSFM}, we have shown that a
noncommutative standard inflationary model, in which a homogeneous
scalar field minimally coupled to gravity, can be a successful
model during the inflation.
\section*{Acknowledgements}
\indent

We thank the Research Council of Shahid Beheshti University for
financial support.

%
\end{document}